\documentclass[aps,prd,preprintnumbers,superscriptaddress,nofootinbib,twocolumn]{revtex4}%
\usepackage{bm,latexsym,amsmath,amssymb,amsfonts,mathrsfs}
\input{colordvi.tex}
\usepackage {cancel}
\usepackage{hyperref}
\usepackage{url}
\usepackage{ulem}
\hypersetup{
    colorlinks=true,
    citecolor=cyan,
}

\newcommand*{\dd}{\text{\large$\cdot$}}
\begin{document}

\title{Effective Field Theory of Large Scale Structure in modified gravity 
and application to Degenerate Higher-Order Scalar-Tensor theories
}

\author{Shin'ichi Hirano}
\email[Email: ]{hirano.s.ai``at''m.titech.ac.jp}
\affiliation{Department of Physics, Tokyo Institute of Technology, 2-12-1 Ookayama, Meguro-ku, Tokyo 152-8551, Japan}

\author{Tomohiro Fujita}
\email[Email: ]{tomofuji``at''aoni.waseda.jp}
\affiliation{Waseda Institute for Advanced Study, Waseda University, Shinjuku, Tokyo 169-8050, Japan}
\affiliation{Research Center for the Early Universe,  The University of Tokyo, Bunkyo, Tokyo 113-0033, Japan}

\begin{abstract}
In modified gravity, the one-loop matter power spectrum exhibits an ultraviolet divergence as shown in the framework of the degenerate higher-order scalar-tensor theory. 
To address this problem, we extend the effective field theory of large scale structure to modified gravity theories. We find that new counterterms appear and renormalize the ultraviolet divergence as a natural consequence of non-linearity in the modified Poisson equation. The renormalized one-loop matter power spectrum is useful to test modified gravity theories by comparing to observations. 
\end{abstract}
\maketitle

\section{Introduction}\label{sec: introduction}

The origin of the accelerating expansion of the Universe is one of the outstanding puzzles of modern physics.
Modified gravity is an interesting candidate for its origin without the cosmological constant.
From this perspective, it is important to test modifications of gravity on cosmological scales by comparing theoretical predictions and observations.

The extensions of general relativity (GR) have various directions.
One of the simplest possibilities is adding a scalar degree of freedom to GR outside the Lovelock theorem~\cite{Lovelock:1971yv} (see recent review~\cite{Langlois:2018dxi,Kobayashi:2019hrl}). 
Such scalar-tensor theories have been intensively investigated.
A typical model is the Horndeski theory~\cite{Horndeski:1974wa,Deffayet:2011gz,Kobayashi:2011nu}.
The Horndeski theory is the most general scalar-tensor theory with second-order equations of motion with respect to the scalar field and the metric tensor. 
Its extension is the Degenerate Higher-Order Scalar-Tensor theory~\cite{Langlois:2015cwa,Crisostomi:2016czh,BenAchour:2016cay} (DHOST), whose Euler-Lagrange equations include higher-order derivatives, while its system keeps second order.
Further extensions have been studied in Refs.~\cite{DeFelice:2018ewo,DeFelice:2021hps,DeFelice:2022xvq,Takahashi:2022mew,Naruko:2022vuh}.
The Horndeski theory and DHOST theory have non-linear derivative scalar self-interactions. Thanks to these interactions, so-called k-mouflage/kinetic screening~\cite{Babichev:2009ee} or Vainshtein screening~\cite{Vainshtein:1972sx} works, \textit{i.e.},
the scalar field does not propagate on small scales, and Newtonian gravity is recovered around the solar system. 

Large scale structure (LSS) is useful for testing modified gravity theories on cosmological scales.
Many galaxy survey projects such as DESI, PFS, and Euclid are ongoing and upcoming~\cite{DESI:2016fyo,PFSTeam:2012fqu,Amendola:2016saw}. These new observations will probe the properties of dark energy and soon achieve, for instance, a percent-level accuracy of its equation of state parameter.
Provided that unprecedented amount of observational data with greatest accuracy will be available, it is crucial to make strict theoretical predictions of observables in modified gravity models and clarifies the differences from those in GR. 
In GR, the evolution of the density fluctuation is determined by the Einstein and fluid equations (continuity and Euler equations) inside the horizon.
In modified gravity, however, the Einstein equations are modified at both linear and non-linear levels.
In the DHOST theory, the linear growth of the density fluctuation is modified~\cite{Kobayashi:2014ida,Tsujikawa:2015mga,DAmico:2016ntq,Crisostomi:2017pjs,Hirano:2019nkz}, and higher-order correlation functions have different signatures from those in GR and the Horndeski theory~\cite{Cusin:2017wjg,Hirano:2018uar,Crisostomi:2019vhj,Lewandowski:2019txi,Yamauchi:2021nxw}. 
It is particularly remarkable that the one-loop correction of the matter power spectrum has a logarithmic ultraviolet (UV) divergence in the DHOST theory according to a perturbative calculation~\cite{Hirano:2020dom}.
If this divergent matter power spectrum was a solid prediction, some parts of the DHOST theory
would be excluded by a serious tension with the non-divergent observed power spectrum. 

Despite its success on large scales, the perturbative treatment of the density fluctuation is broken down on small scales $(\lesssim {\cal O}({\rm Mpc}))$ because its perturbative quantity exceeds unity there.
It is known that the perturbative expansion is broken down at three-loop even in GR~\cite{Blas:2013aba}.
Toward valid predictions, the effective field theory of large scale structure (EFTofLSS)~\cite{Baumann:2010tm,Carrasco:2012cv,Hertzberg:2012qn} has been developed.
The EFTofLSS takes into account the effects of small-scale physics in a fashion inspired by effective field theory approach. 
The effective fluid is introduced to incorporate the small-scale physics, and it provides counterterms and eliminates a problematic UV sensitivity, which otherwise picks up the breakdown of the perturbative approach.
As a result, the perturbative expansion is converged.
As a price to pay, the so-called EFT parameters appear, and it is necessary to fix them by observations or simulations.

In this paper, we extend the EFTofLSS to modified gravity and demonstrate our framework in the DHOST theory. 
The essence is that new counterterms appear in modified gravity and they renormalize the UV divergence.
In the EFTofLSS in GR, non-linear interactions exist only in the Euler equation and the counterterms arise from them.
In modified gravity, however, non-linear interactions also exist in the equations of motion for the scalar field and the gravity potentials. 
Thus we need to introduce new effective terms in the modified Poisson equation, which leads to novel counterterms.
As we show later, these new counterterms exactly cancel the UV divergence in the one-loop corrections to the matter power spectrum and the prediction of the DHOST theory becomes convergent.

This paper is organized as follows. In Sec~\ref{sec: EFTofLSS}, we shortly review the EFTofLSS in GR and illustrate that the appropriate counterterms appear from the effective fluid induced by non-linear interactions in the Euler equation.  
In Sec.~\ref{sec: EFTofLSSinMG}, we extend the framework of the EFTofLSS in GR to modified gravity theories and show that the appropriate new counterterms appear from the non-linearity in the modified Poisson equation. Applying it to the DHOST theory, we also demonstrate the cancellation of the UV divergence in the one-loop power spectrum.
Sec.~\ref{sec: conclusion} is devoted to our conclusion.

\section{Review of EFTofLSS in GR}\label{sec: EFTofLSS}

In this section, we briefly review the EFTofLSS in the case of GR, before extending it in the next section. We refer the interested reader to a recent review and the references therein~\cite{Cabass:2022avo}. 

\subsection{EFT of long modes}

Let us consider the dynamics of the density fluctuation of the matter field inside horizon.
The metric perturbation around a spatially flat homogeneous and isotropic universe is defined by
\begin{align}
    ds^2 &= -[1 +2\Phi(t,{\bm x})]dt^2 +a^2(t)[1 -2\Psi(t,{\bm x})]d{\bm x}^2.
\end{align}
The fluctuations of the matter field is characterized by the density fluctuation $\delta$ and the scalar component of the velocity field $\theta$ defined as
\begin{align}
    \delta(t,{\bm x}) \equiv  \frac{\rho(t,{\bm x})-\bar\rho(t)}{\bar\rho(t)},
\qquad
    \theta(t,{\bm x}) \equiv  \frac{\partial_{i}v^{i}( t,{\bm x})}{aH},\label{eq: theta}
\end{align}
where $\rho$ is the energy density of the matter, $\bar\rho$ is its mean value, the velocity field of the matter denotes $v^{i}$, and we ignore its vectotial component.
In GR, the dynamics of gravitational potentials are given by Einstein equations inside the cosmological horizon, $i.e.,$ Poisson equation. $\delta$ and $\theta$ are governed by the continuity and Euler equations.
These equations form a closed, coupled and non-linear system,
\begin{align}
    &\partial^2\Phi = \frac{3}{2}a^2H^2\Omega_{\rm m}\delta, \label{eq: PoissonR}
\\
    &\dot\delta +\frac{1}{a}\partial_i[(1+\delta)v^i] =0,\label{eq: continuityR}
\\
    &\dot v^i +Hv^i +\frac{1}{a}v^j\partial_{j}v^{i}+\frac{1}{a}\partial^i\Phi =0,\label{eq: EulerR}
\end{align}
where $H=\dot a/a$ and $\Omega_{\rm m}= 8\pi G_{\rm N}\bar\rho/3H^2$. An overdot denotes the derivative with respective to the time coordinate, and $G_{\rm N}$ is the gravitational constant.
It should be stressed that the gravitational equation is still Poisson equation even if we consider the non-linear density fluctuation. Non-linearity arises only from the Euler equation in the large scale structure in GR. 

In the so-called standard perturbation theory (SPT), one solves Eqs.~\eqref{eq: PoissonR}-\eqref{eq: EulerR} in perturbation with respect to $\delta$ and $\theta$ in Fourier space~\cite{Bernardeau:2001qr}. On small scales, however, these perturbative variables exceed unity and the perturbation theory breaks down, which propagates to larger scales through loop corrections and hinders the correct convergence~\cite{Blas:2013aba}. To address this problem, the EFTofLSS has been developed~\cite{Baumann:2010tm,Carrasco:2012cv,Hertzberg:2012qn}.
The original perturbative variables are split into short mode and long mode as
\begin{align}
    \delta(t,{\bm x}) &= \delta_{s}(t,{\bm x} ) +\delta_{l}(t,{\bm x} ),
\\
    v^i(t,{\bm x}) &= v^i_{s}(t,{\bm x} ) +v^i_{l}(t,{\bm x} ),
\\
    \Phi(t,{\bm x}) &= \Phi_{s}(t,{\bm x} ) +\Phi_{l}(t,{\bm x} ),
\end{align}
where the long-mode variables are defined by a smoothing with an appropriate window function $W_{\Lambda}$,
\begin{align}
    X_{l}(t,{\bm x} ) &= \int d^3x'~W_{\Lambda}(|{\bm x}-{\bm x}'|)X(t,{\bm x}'),
    \label{eq: Xell}
\end{align}
with $X=\delta, \Phi$ and $\rho v^i$.
Here, $\Lambda$ is a coarse-grained (cutoff) scale with a hierarchy, $H\ll \partial/a\ll \Lambda$. 

The EFTofLSS describes the dynamics of the smoothed long-mode variables $X_l$ which are sufficiently small and hence enable a valid perturbation. The short modes affect the long-mode dynamics only through the non-linear term in the Euler equation~(\ref{eq: EulerR})~\cite{Baumann:2010tm},
\begin{align}
    v^j\partial_{j}v^{i} = v_l^j\partial_{j}v_l^{i} +\frac{1}{\rho_l}\partial_{j}\tau^{ij}_{\Lambda},\label{eq: non-l term}
\end{align}
where $\tau^{ij}_{\Lambda}$ is the energy momentum tensor of the short modes,
\begin{align}
    \tau^{ij}_{\Lambda} &= \rho_{l} v_{s}^iv_{s}^j
    -\frac{1}{8\pi G_{\rm N}}[(\partial^k\Phi_s)^2\delta^{ij} -2\partial^i\Phi_s\partial^j\Phi_s].
\label{eq: tau}
\end{align}
Here higher order derivative terms ${\cal O}\left(\partial^2/(a\Lambda)^2\right)$ have been ignored.
The EFTofLSS rewrites $\tau^{ij}_{\Lambda}$ in terms of the long modes not by solving the short modes but by introducing an effective fluid expression,\footnote{$\langle...\rangle_{\delta_l}$ denotes the ensemble average over the short modes under the influence of the long wavelength background $\delta_l$. This procedure corresponds to writing down all possible terms in the EFT of QFT. }
\begin{align}
   \rho^{-1}_{l}\langle \tau^{ij}_{\Lambda}\rangle_{\delta_{l}} &= 
   \left[c_{\rm s}^2\delta_l -\frac{c_{bv}^2}{a H}\partial_kv_l^k\right]\delta^{ij}
   +\Delta\tau^{ij}
\notag\\   
   &\quad -\frac{3}{4}\frac{c_{ sv}^2}{a H}\left(\partial^iv_l^j+\partial^jv_l^i -\frac{2}{3}\partial_kv_l^k\delta^{ij}\right),
   \label{eq: EFT fluid}
\end{align}
which includes all possible linear terms of the long modes that respect the symmetry $i\leftrightarrow j$. The coefficients $c_i~(i={s},{cv},{sv})$ are undetermined functions of time and the cutoff scale at this point.
$\Delta\tau$ is the stochastic term representing the stochastic noise of short modes to long modes~\cite{Peebles:1980co}.
For the above EFT terms to be significant from the one-loop corrections, their perturbative order are assigned as $c_i = {\cal O}(\delta_l^2)~(i={s},{cv},{sv})$ and $\Delta\tau = {\cal O}(\delta_l^3)$.

\subsection{Proper counterterms}

In what follows, we shall see that the effective fluid terms \eqref{eq: EFT fluid} lead to additional contributions which cancel the UV sensitivity of the SPT solution. 
We send the detailed calculations to App.~\ref{sec: EFT GR app} and concisely explain essential points.
Putting the above equations altogether, one obtains a closed system of equations for the long modes (see Eqs.~\eqref{eq: continuityEFT} and \eqref{eq: EulerEFT}).
To solve this system perturbatively, we expand the long modes as
\begin{align}
    \Phi_{l} = \sum_{n=1}\Phi_{l}^{(n)},
    \quad
    \delta_{l} = \sum_{n=1}\delta_{l}^{(n)},\quad
    \theta_{l} = \sum_{n=1}\theta_{l}^{(n)},
\end{align}
where $\Phi_{l}^{(n)}, \delta_{l}^{(n)}, \theta_{l}^{(n)}$ are $n$-th order quantities in $\delta_{l}^{(1)}$.
Substituting them into the coupled equations, we can determine $n$-th order solutions order by order.
Up to second order, the solutions are the same as the SPT. One finds the linear solution as
\begin{align}
    \delta_{l}^{(1)}(t,{\bm p}) &= D_{+}(t)\delta_{\rm{L}}({\bm p}),\label{eq: linear soln}
\end{align}
where $\delta_{\rm{L}}({\bm p})$ is the initial density field and $D_{+}$ is the growth factor.
The second-order solution is given by
\begin{align}
    \delta^{(2)}_{l}(t,{\bm p}) &=
    \frac{D_+^2(t)}{(2\pi)^3}\int^{\Lambda} d^3k~ F_2({\bm p},{\bm k})\delta_{l}({\bm p}-{\bm k})\delta_{l}({\bm k}),
\end{align}
where the standard second-order kernel $F_2$ can be found in Eq.~\eqref{eq: F2GR}.
The third-order solution acquires additional contributions from the effective fluid~\eqref{eq: EFT fluid},
\begin{align}
     \delta^{(3)}_{l}(t ,{\bm p}) &=  \delta^{(3)}_{\rm SPT}(t ,{\bm p}) + \delta^{(3)}_{{\rm EFT}1}(t ,{\bm p})+\delta^{(3)}_{{\rm EFT}2}(t ,{\bm p}),
\end{align}
where the second term in the right hand side, which arises from the terms with $c_i$ in Eq.~\eqref{eq: EFT fluid}, reads
\begin{align}
\delta^{(3)}_{\rm EFT1}(t ,{\bm p}) &= -p^2\int^{t}d\tau\, G(t,\tau)c_{\rm comb}^2H^2\delta_l^{(1)}(\tau,{\bm p}),
\label{eq: EFT3rd sol1}
\end{align}
Here, $G$ is the Green function in Eq.~(\ref{eq: evolution2}) and $c_{\rm comb}^2 \equiv c_{s}^2+(c_{sv}^{2}+c_{bv}^{2})f$ is a new coefficient.
$\delta^{(3)}_{\rm EFT2}$ and $\delta^{(3)}_{\rm SPT}$ are given in Eqs.~\eqref{eq: delEFT2} and \eqref{eq: delSPT3}, respectively.

The one-loop matter power spectrum is written as
\begin{align}
    &P_{1{\rm loop}}(t,p) = P_{11}(t,p) +P_{22}(t,p) +2P_{13}(t,p),
\end{align}
with
\begin{align}
    & \langle \delta^{(1)}(t,{\bm p})\delta^{(1)}(t,{\bm p}') \rangle 
        = (2\pi)^{3}\delta_{\rm D}({\bm p}+{\bm p}')P_{11}(t,p), 
        \label{eq: P11}
\\
    & \langle \delta^{(2)}(t,{\bm p})\delta^{(2)}(t,{\bm p}') \rangle 
        = (2\pi)^{3}\delta_{\rm D}({\bm p}+{\bm p}')P_{22}(t,p), 
        \label{eq: P22}
\\
    & \langle \delta^{(1)}(t,{\bm p})\delta^{(3)}(t,{\bm p}') \rangle 
        = (2\pi)^{3}\delta_{\rm D}({\bm p}+{\bm p}')P_{13}(t,p).
        \label{eq: P13}
\end{align}
The linear power spectrum $P_{11}(t,p)= D_{+}(t)^2P_{\rm L}(p)$ is related to 
the initial linear power spectrum $P_L(p)$ defined by $\langle \delta_{\rm{L}}({\bm p})\delta_{\rm{L}}({\bm p}')\rangle = (2\pi)^3\delta_{\rm D}({\bm p}+{\bm p}')P_{\rm L}(p)$.

We focus on the following two contributions which come from $\langle \delta^{(1)}\delta^{(3)}_{\rm SPT} \rangle$ and $\langle \delta^{(1)}\delta^{(3)}_{\rm EFT1} \rangle$, respectively:
\begin{align}
P_{13}^{\rm SPT}(t,p) &= \frac{2D_+^4(t)}{(2\pi)^3}P_{L}(p)\int^{\Lambda} d^3k~ F_3({\bm p},{\bm k},-{\bm k})P_{L}(k),
\label{eq: 13SPT}
\\
P_{13}^{\rm EFT1}(t,p) &= 
C_{\rm comb}(t, \Lambda)\, p^2P_{L}(p),
 \label{eq: P13 EFT1}
\end{align}
where $F_3$ can be found in Eq.~\eqref{eq: F3GR} and $C_{\rm comb}(t,\Lambda)\equiv -D_+(t)\int^{t}d\tau G(t,\tau)c_{\rm comb}^2H^2 D_+(\tau)$ is the final form of the undetermined coefficient.
$C_{\rm comb}$ depends on the cutoff scale $\Lambda$, which characterizes the long mode through Eq.~\eqref{eq: Xell}.
After all, Eq.~\eqref{eq: P13 EFT1} is the contribution from the EFT terms with $c_i$ in Eq.~\eqref{eq: EFT fluid} to the power spectrum that is $p^2 P_L(p)$ multiplied by an undetermined time-dependent function.

In the SPT, $P_{13}^{\rm SPT}$ has a UV sensitivity (i.e. dependence on short scale physics), because the upper limit of the integral is originally infinity, although it is replaced by the cutoff scale $\Lambda$ in the EFTofLSS. The loop integral can potentially pick up the breakdown of the perturbation theory on short scales. Nevertheless, in the UV limit of the loop integral, $P_{13}^{\rm SPT}$ behaves as
\begin{align}
    P^{\rm SPT}_{13} &\approx -\frac{61D_+^4(t)}{315(2\pi)^2}p^{2}P_{\rm L}(p)
\int^{\Lambda}_{p\ll k} dk ~P_{\rm L}(k).
\label{eq: SPT13 UV}
\end{align}
For concreteness, the Einstein--de Sitter universe is assumed only in Eq.~\eqref{eq: SPT13 UV}.
We find $P^{\rm SPT}_{13}\propto p^2 P_L(p)$ and it has the same dependence on the external momentum as the EFT contribution $P_{13}^{\rm EFT1}$. The potential UV sensitivity of $P^{\rm SPT}_{13}$ can be cancelled by $P_{13}^{\rm EFT1}$ using the freedom of its undetermined coefficient $C_{\rm comb}$. In other words, $P_{13}^{\rm EFT1}$ works as a proper counterterm with the correct momentum dependence. 
It is also shown in App.~\ref{sec: EFT GR app} that $\delta^{(3)}_{{\rm EFT}2}$ has the same momentum dependence as the UV limit of $P_{22}^{\rm SPT}$ and hence serves as another proper counterterm.

Fortunately, since the standard linear power spectrum quickly decays on short scales $P_{\rm L}\propto k^{-3}~(k\gg k_{\rm eq})$, the loop integrals of $P^{\rm SPT}_{13}$ and $P^{\rm SPT}_{22}$ do not possess significant UV sensitivity in our Universe. 
However, if the decay is slower $P_{\rm L}\propto k^n~(n>-1)$, the loop integral of $P^{\rm SPT}_{13}$ exhibits UV divergence even in GR. 
Even in such an extreme case, the leading UV divergence from $P^{\rm SPT}_{13}$ can be renormalized by 
$P_{13}^{\rm EFT1}$.

\subsection{Renormalization and fixing EFT parameter}\label{sec: fitting GR}

The renormarlized one-loop power spectrum in the EFTofLSS in GR is given by
\begin{equation}
    P^{\rm EFT}_{\rm 1loop}= P_{11}+P_{22}^{\rm SPT}+2P_{13}^{\rm SPT}+2P_{13}^{\rm EFT1},
\end{equation}
where the undertermined parameter $C_{\rm comb}$ is still included in the last term~\eqref{eq: P13 EFT1} and we ignored $P_{13}^{\rm EFT2}$ as a small contribution. 
To fix the EFT parameter $C_{\rm comb}$, we need to use a reference value $P_{\rm obs}$ at a certain wavenumer $p_{\rm ren}(<\Lambda)$ from observation or simulation as
\begin{equation}
    C_{\rm comb}(t,\Lambda)= \left.\frac{P_{\rm obs}-P_{11}-P_{22}^{\rm SPT}-2P_{13}^{\rm SPT}}{2p^2P_{L}}\right|_{p=p_{\rm ren}}.
\end{equation}
Plugging it into $P^{\rm EFT}_{\rm 1loop}$, we can make a prediction of the matter power spectrum for $p\neq p_{\rm ren}$.
Note that the above equation also gives the running of $C_{\rm comb}$ as the renormalization scale $p_{\rm ren}$ changes.

One may wonder how the UV sensitivity of $P_{13}^{\rm SPT}$ was cancelled in the above treatment. To explicitly see that,
we divide the coefficient $C_{\rm comb}$
into two parts,
\begin{align}
    C_{\rm comb}(t,\Lambda) &=
     C_{\rm ren}(t)
     +C_{\rm ctr}(t,\Lambda).
\end{align}
The second term $C_{\rm ctr}$ cancels the leading UV behavior of  $P^{\rm SPT}_{13}$, and the first term remains as the renormalized part of $P^{\rm EFT1}_{13}$. Using Eq.~\eqref{eq: SPT13 UV}, the concellation condition is written as
\begin{align}
    &C_{\rm ctr}(t,\Lambda)-\frac{61D_+^4(t)}{315(2\pi)^2}\int^{\Lambda}_{p\ll k} dp ~P_{\rm L}(k) =0.
\end{align}
Under this condition, the leading $\Lambda$ dependence vanishes from $P^{\rm EFT}_{\rm 1loop}$.

\section{EFTofLSS in modified gravity}\label{sec: EFTofLSSinMG}

Now we consider a modification of gravity theory from GR.
A typical example of modified gravity is a scalar-tensor theory.
The Horndeski theory~\cite{Horndeski:1974wa,Deffayet:2011gz,Kobayashi:2011nu} is the most general scalar-tensor theory whose equations of motion are second order. 
As a further extension, the Degenerate Higher-Order Scalar-Tensor (DHOST) theory~\cite{Langlois:2015cwa,Crisostomi:2016czh,BenAchour:2016cay} is known as a more general theory including higher derivative operators, but its system is still second order. 
The DHOST theory provides a powerful framework including many concrete models of scalar-tensor theories.

The one-loop matter power spectrum in the DHOST theory has been derived
in the SPT manner~\cite{Hirano:2020dom}. Interestingly, it was pointed out that the one-loop power spectrum has the logarithmic divergence even with the standard linear power spectrum. 
In the present paper, we develop the EFTofLSS in modified gravity 
and show that new counterterms cancelling the logarithmic divergence
arises as a natural consequence of novel non-linear dynamics introduced by the modification of the Poisson equation.

\subsection{SPT in modified gravity}\label{sec: SPTinMG}

We quickly review the results of the SPT calculations in the DHOST theory.
The linear evolution equation of the density contrast is modified as~\cite{Hirano:2019nkz}
\begin{align}
	&\ddot\delta^{(1)}_{l} + [2+\varsigma(t)]H\dot\delta^{(1)}_{l} -\frac{3}{2}\Omega_{\mathrm{m}}\Xi_{\Phi}(t)H^{2}\delta^{(1)}_{l} = 0.
	\label{eq: evolution1inMG}
\end{align}
$\varsigma$ and $\Xi_{\Phi}$ represent the effects of the modification of the gravity (in GR, $\varsigma=0$ and $\Xi_{\Phi}=1$). 
Within the Horndeski theory, $\varsigma=0$ and $\Xi_{\Phi}\neq1$. 
In the DHOST theory, $\varsigma\neq0$ and $\Xi_{\Phi}\neq1$.
While the coefficients of the evolution equation are different from those in GR, the form of the linear solution is same as that in GR, Eq.~(\ref{eq: linear soln}).
The linear effect of modification of gravity is encoded into the growth factor, $D_+$.

In the DHOST theory, the kernel functions in the second and third-order solutions are generalized as
\begin{align}
	&F_{2}(t,{\bm p}_{1},{\bm p}_{2}) = \kappa(t)\,  \alpha_s({\bm p}_{1},{\bm p}_{2}) -\frac{2}{7}\lambda(t)\,  \gamma({\bm p}_{1},{\bm p}_{2}),
	\label{eq: F2MG}
\\
	&F_{3}(t,{\bm p}_{1},{\bm p}_{2},{\bm p}_{3}) = d_{\alpha\alpha}(t)\, \alpha\alpha ({\bm p}_{1},{\bm p}_{2},{\bm p}_{3}) 
	\label{eq: F3MG}
\\
	&\quad -\frac{4}{7}d_{\alpha\gamma }(t)\, \alpha\gamma ({\bm p}_{1},{\bm p}_{2},{\bm p}_{3})
	-\frac{2}{21}d_{\gamma\gamma}(t)\, \gamma\gamma({\bm p}_{1},{\bm p}_{2},{\bm p}_{3})
\notag\\
	&\quad	
	+\frac{1}{9}d_{\xi}(t)\, \xi_c({\bm p}_{1},{\bm p}_{2},{\bm p}_{3})
	+d_{\alpha\alpha\ominus}(t)\, \alpha\alpha_{\ominus}({\bm p}_{1},{\bm p}_{2},{\bm p}_{3}) 
\notag\\
	&\quad +d_{\alpha\gamma\ominus}(t)\, \alpha\gamma_{\ominus}({\bm p}_{1},{\bm p}_{2},{\bm p}_{3})
	+d_{\zeta}(t)\, \zeta_c({\bm p}_{1},{\bm p}_{2},{\bm p}_{3}).
\notag
\end{align}
The explicit formulae of these third-order shape functions, $\alpha\alpha_{\ominus},~\alpha\gamma_{\ominus},~\zeta_c$, are given in App~\ref{sec: 3rd shape funcs}.
The modification of gravity is encoded into the coefficients, $\kappa$, $\lambda$, and $d$s (these explicit forms in the DHOST theory were shown in~\cite{Hirano:2020dom}).
In GR, $\kappa=\lambda=d_{\alpha\alpha}=d_{\alpha\gamma}=d_{\gamma\gamma}=\xi_c=1$ and $d_{\alpha\alpha\ominus}=d_{\alpha\gamma\ominus}=d_{\zeta}=0$~(see Eq.~(\ref{eq: F2GR}) and Eq.~(\ref{eq: F3GR})).
Within the Horndeski theory, $\kappa=d_{\alpha\alpha}=1$, $\lambda$, $d_{\alpha\gamma}$, $d_{\gamma\gamma}$, $d_{\xi}$ can deviate from unity, and $d_{\alpha\alpha\ominus}=d_{\alpha\gamma\ominus}=d_{\zeta}=0$.
In the DHOST theory, all coefficients can deviate from standard ones, in particular, $d_{\alpha\alpha\ominus}$, $d_{\alpha\gamma\ominus}$, and $d_{\zeta}$ can take non-zero values.

Using these kernel functions,
one obtains the one-loop corrections to the matter power spectrum.
In the UV limit, the leading behaviour of the one-loop corrections are~\cite{Hirano:2020dom}
\begin{align}
    P^{\rm SPT}_{13}(t,p)
	&\approx Q_{13}(t)\, P_{\rm L}(p)
	\int^{\Lambda}_{p\ll k}{\rm d}k\, k^2\,
	P_{\rm L}(k)
	\notag\\
	&\quad+ \tilde{Q}_{13}(t)\, p^2 P_{\rm L}(p) 
	\int^{\Lambda}_{p\ll k}{\rm d}k\,
	P_{\rm L}(k),
	\label{eq: P13MG}
\\
	P^{\rm SPT}_{22}(t,p)
	&\approx Q_{22}(t)\, p^4 
	\int^{\Lambda}_{p\ll k} {\rm d}k\, k^{-2}\, P^2_{\rm L}(k),
\end{align}
where $Q_{13}\equiv -D_{+}^4\left( d_{\alpha\alpha\ominus}+d_\zeta\right)/(6\pi^2)$, 
$\tilde{Q}_{13}\equiv D_{+}^4$ $\times[147d_{\alpha\alpha}-144d_{\alpha\gamma}-64d_{\gamma\gamma} -357d_{\alpha\alpha\ominus}-252d_{\alpha\gamma\ominus}-210d_{\zeta}]/(1260\pi^2)$,
and $Q_{22}\equiv D_{+}^4(343\kappa^2-336\kappa\lambda+128\lambda^2)/(2940\pi^2)$.
Note that $Q_{13}$ vanishes and $\tilde Q_{13}$ becomes the leading term in the Horndeski. Furthermore, the $\tilde Q_{13}$ term reproduces Eq.~\eqref{eq: 13SPT} in the GR limit.
The momentum dependence of $P^{\rm SPT}_{22}$ remains the same as the GR case (see Eq.~\eqref{eq: P22SPT}).
However, the integrand of the $Q_{13}$ term gains an extra factor of $k^2$ compared to the GR case in Eq.~\eqref{eq: SPT13 UV}.
Since the standard linear power spectrum is $P_{\rm L}\propto k^{-3}$ in the UV regime, this loop integral leads to a logarithmic divergence~\cite{Hirano:2020dom}. Thus the matter power spectrum in the DHOST theory exhibits a serious UV sensitivity.
To cancel this stronger UV sensitivity, we need an additional counterterm on top of the effective fluid terms in the previous section, because the dependence on the external momentum is also changed
from Eq.~\eqref{eq: SPT13 UV}.

\subsection{New counterterms}

It is important to remember that the counterterms arises from the non-linear interactions
in the Euler equation~\eqref{eq: non-l term} in the EFTofLSS in GR. 
This is because the short modes can affect the long mode dynamics only through non-linear interactions. 
Of course, one can consider the EFT terms caused by the same origin in 
dark energy and modified gravity models~\cite{Cusin:2017wjg}.
However, in modified gravity, there exist the other non-linear interactions which induce new counterterms.

In scalar-tensor modified gravity, the scalar field and gravitational potentials interact at not only linear level but also non-linear level in contrast to GR. 
As a result, the Poisson equation changes and includes non-linear interactions of gravitational potentials as well as the modified gravitational constant and the extra friction term.
Schematically, the modified Poisson equation can be written as (see App.~\ref{sec: schematic Poisson} for derivation)
\begin{align}
  \frac{1}{a^2H^2}\partial^2\Phi
  +\frac{1}{a^4H^4}\mathcal{T}_{\rm NL} 
  = \mu_\Phi \delta + \frac{\nu_\Phi}{H}\dot{\delta}+\frac{\kappa_\Phi}{H^2}\ddot{\delta},
  \label{eq: modified Poisson}    
\end{align}
with non-linear terms
\begin{align}
\mathcal{T}_{\rm NL}=& 
    \tilde\tau^{(2)}_{\Phi,\alpha}(t) \left[(\partial^2\Phi)^2+\partial_{i}\Phi\partial_{i}\partial^2\Phi\right]
\notag\\
  &+
   \tilde\tau^{(2)}_{\Phi,\gamma}(t) \left[(\partial^2\Phi)^2-(\partial_{i}\partial_{j}\Phi)^2\right],
\end{align}
where the coefficients, $\mu_\Phi,\nu_\Phi, \kappa_\Phi, \tilde\tau$s, are written by the growth rate and the background variables. 
$\mu_\Phi$, $\nu_\Phi$, and $\tilde{\tau}^{(2)}_{\Phi,\alpha}$ vanish in the Horndeski theory while these can be non-zero in the DHOST theory.
The higher-order terms $\mathcal{O}(\Phi^3)$ are neglected.
Smoothing out the above equation, these non-linear interactions yields quadratic terms of the short modes in the same way as Eq.~(\ref{eq: non-l term}). 
Following the procedure of the EFTofLSS in GR (see Eq.~\eqref{eq: EFT fluid}), 
in rewriting the non-linear terms of the short modes in terms of the long mode,
we consider all possible linear term of $\Phi_l$ as an effective description,
\begin{align}
\frac{\langle \mathcal{T}_{\rm NL}\rangle_{\delta_{l}}}{a^4H^4}=
  c_{1}\Phi_l+
  c_{2}\frac{\partial^2}{a^2H^2}\Phi_l+
  c_{3}\frac{\partial^2}{a^2H^2}\frac{\partial^2}{a^2\Lambda^2}\Phi_l,
  \label{eq: EFT terms}
\end{align}
where higher derivative terms are omitted.
We regard these terms as ${\cal O}(\delta_{l}^3)$ such that they contribute to the one-loop power spectrum.
We expect that $c_1, c_2$, and $c_3$ are small parameters including power laws of $H/\Lambda$ in the analogy of EFT in quantum field theory (i.e. integrating out heavy degrees of freedom, then one obtain counterterms for couplings).

In the perturbative analysis of the DHOST theory, the quasi-static approximation is applied in which the time dependence of both background variables and perturbations are assumed to be the order of the Hubble parameter and hence negligible for the sub-horizon dynamics.  Then the gravitational potential always appears as $\partial^2\Phi$ in the modified Poisson equation accompanied by squared spatial derivative.
In this paper, therefore, we exclude the $c_1$ term and consider only the $c_2$ and $c_3$ terms as the leading corrections. Note that an important exception of the above argument is the Chameleon gravity~\cite{Brax:2004qh}.
In the Chameleon case, the mass of the scalar field $m$ provides a relevant time scale for background variables, and the time derivatives are not necessarily negligible. Indeed, its modified Poisson equation is known to acquire a mass term, $m^2\Phi$.
Thus, the $c_1$ term would be included as a correction term to the mass term. 

The EFT terms~\eqref{eq: EFT terms} are regarded as source terms in the evolution equation for 
$\delta^{(3)}$.
Using the linear solution~\eqref{eq: linear soln} and modified Poisson equation with the EFT terms (\ref{eq: modified Poisson}) and (\ref{eq: EFT terms}), we obtain the third-order evolution equations
\begin{align}
    &\ddot\delta^{(3)}_{l} + [2+\varsigma(t)]H\dot\delta^{(3)}_{l} -\frac{3}{2}\Omega_{\mathrm{m}}\Xi_{\Phi}(t)H^{2}\delta^{(3)}_{l} 
\notag\\
    &= \left(\tilde c_{2} -\tilde c_{3}\frac{p^{2}}{a^2\Lambda^2}\right)\delta^{(1)}_{l} +({\rm SPT~ parts}),
	\label{eq: evolution3rdEFT}
\end{align}
where
\begin{align}
    \tilde c_2 &= 
    c_2 \frac{\kappa_\Phi 
    +\nu_\Phi f 
    +\mu_\Phi\left[(fH)^{\dd}/H^2 +f^2\right]}{1-\mu_\Phi},
\\
    \tilde c_3 &= 
    c_3 \frac{\kappa_\Phi 
    +\nu_\Phi f 
    +\mu_\Phi\left[(fH)^{\dd}/H^2 +f^2\right]}{1-\mu_\Phi}.
\end{align}

The contributions to the third-order solution from the EFT corrections read
\begin{align}
    \delta^{(3)}_{c_2}(t,p) &= 
    \int^{t}d\tau\, \tilde G(t,\tau)\, 
    \tilde{c}_2(\tau)
    \delta^{(1)}_l(\tau,p),
\\
    \delta^{(3)}_{c_3}(t,p) &= 
    -\frac{p^2}{\Lambda^2}\int^{t}d\tau \frac{ \tilde G(t,\tau) }{a^2(\tau)}\,
    \tilde{c}_3(\tau)
    \delta^{(1)}_l(\tau,p).
\end{align}
where $\tilde G$ is the Green function of Eq.~(\ref{eq: evolution3rdEFT}). 
The one-loop corrections from these counterterms are
\begin{align}
   P^{c_2}_{13}(t,p) &= P_{\rm L}(p)D_{+}(t)\int^{t}d\tau \tilde G(t,\tau)
   \tilde{c}_2(\tau)
   D_{+}(\tau), 
\notag\\
   &\equiv C_{2}(t,\Lambda) P_{\rm L}(p),\label{eq: c2}
\\[2mm]
    P^{c_3}_{13}(t,p) &= 
    -\frac{p^2}{\Lambda^2}P_{\rm L}(p)
D_{+}(t)\int^{t}d\tau \frac{\tilde G(t,\tau)}{a^2(\tau)}\tilde{c}_3(\tau)
    D_{+}(\tau),
\notag\\    
    &\equiv C_{3}(t,\Lambda) p^2P_{\rm L}(p).\label{eq: c3}
\end{align}
These momentum dependences correspond to those of the leading and sub-leading terms of $P^{\rm SPT}_{13}$, Eq.~(\ref{eq: P13MG}).
$P^{c_2}_{13}$ plays the role of the counterterm to the leading term in $P^{\rm SPT}_{13}$ while $P^{c_3}_{13}$ does that to the sub-leading term.
Even in modified gravity, the EFT fluid argument for the Euler equation in Sec.~\ref{sec: EFTofLSS} applies and the counter term $P_{13}^{\rm EFT1}$ in Eq.~\eqref{eq: P13 EFT1} emerges~\cite{Cusin:2017wjg}. However, its momentum dependence is the same as $P^{c_3}_{13}$ and thus $C_{\rm comb}$ can be absorbed in $C_3$. Therefore, we have two undertemined coefficients, $C_2$ and $C_3$, in the DHOST case.

\subsection{Renormalized power spectrum}\label{sec: fitting}

The renoamarlized one-loop power spectrum in the EFTofLSS in the DHOST reads
\begin{equation}
    P^{\rm EFT}_{\rm 1loop}= P_{11}+P_{22}^{\rm SPT}+2P_{13}^{\rm SPT}+2P_{13}^{\rm c_2}+2P_{13}^{\rm c_3},
\end{equation}
where $P_{13}^{\rm SPT}$ is divergent as we saw in Sec.~\ref{sec: SPTinMG} and the EFT parameters, $C_2$ and $C_3$, are included in the last two terms. We ensure the cancellation of the UV sensitivities in the same way as Sec.~\ref{sec: fitting GR}.
Splitting the EFT parameters into two parts, respectively,
\begin{align}
  C_{i}(t,\Lambda) := C_{i}^{\rm ren}(t) +C_{i}^{\rm ctr}(t,\Lambda),
\end{align}
we require the cancellation between the leading and sub-leading terms in Eq.~(\ref{eq: P13MG}) and the counterterms  Eqs.~(\ref{eq: c2}) and (\ref{eq: c3}) as
\begin{align}
    &C^{\rm ctr}_{2}(t,\Lambda)+
    Q_{13}(t)	\int^{\Lambda}_{p\ll k}{\rm d}k\, k^2\, P_{\rm L}(k)=0,
\label{eq:r cond1}
\\
    &C^{\rm ctr}_{3}(t,\Lambda)+
    \tilde{Q}_{13}(t)	\int^{\Lambda}_{p\ll k}{\rm d}k\,  P_{\rm L}(k)=0.
\label{eq:r cond2}
\end{align}
After these cancellations, the divergent $P_{13}^{\rm SPT}$ is renormalized as
\begin{align}
P_{13}^{\rm ren}(p)&\equiv P_{13}^{\rm SPT}(p)- 
Q_{13}(t)\, P_{\rm L}(p)
	\int^{\Lambda}_{p\ll k}{\rm d}k\, k^2\,
	P_{\rm L}(k)
	\notag\\
	&\quad- \tilde{Q}_{13}(t)\, p^2 P_{\rm L}(p) 
	\int^{\Lambda}_{p\ll k}{\rm d}k\, P_{\rm L}(k).
	\label{eq:Pren13}
\end{align}
The loop integral of $P_{13}^{\rm ren}$ is convergent.
With this renormalized $P_{13}$, the one-loop power spectrum is rewritten as
\begin{equation}
    P^{\rm EFT}_{\rm 1loop}= P_{11}+P_{22}^{\rm SPT}+2P_{13}^{\rm ren}+2\tilde{P}_{13}^{\rm c_2}+2\tilde{P}_{13}^{\rm c_3},
\end{equation}
where $\tilde{P}_{13}^{\rm c_i}$ is $P_{13}^{\rm c_i}$ with the replacement of $C_i$ by $C_i^{\rm ren}$.

To fix $C_i^{\rm ren}$, we need reference values $P_{\rm obs}$ at two different wave-numbers, $p^{\rm ren}_1$ and $p^{\rm ren}_2$. Solving
\begin{equation}
    P^{\rm EFT}_{\rm 1loop}(p^{\rm ren}_1)=P_{\rm obs}(p^{\rm ren}_1),
    \quad
    P^{\rm EFT}_{\rm 1loop}(p^{\rm ren}_2)=P_{\rm obs}(p^{\rm ren}_2),
    \label{eq:double rem conditions}\notag
\end{equation}
with respect to $C_2^{\rm ren}$ and $C_3^{\rm ren}$, we obtain
\begin{align}
    C_2^{\rm ren}&=\frac{(p^{\rm ren}_1)^2 P_L^{(1)}
    \Delta P^{(2)}
    -(p^{\rm ren}_2)^2 P_L^{(2)}
    \Delta P^{(1)}
    }{2[(p^{\rm ren}_1)^2-(p^{\rm ren}_2)^2]P_L^{(1)}P_L^{(2)}},
    \\
    C_3^{\rm ren}&=\frac{P_L^{(2)}
    \Delta P^{(1)}
    -P_L^{(1)}\Delta P^{(2)}
    }{2[(p^{\rm ren}_1)^2-(p^{\rm ren}_2)^2]P_L^{(1)}P_L^{(2)}},
    \label{eq:C3ren}
\end{align}
where we defined $P^{(i)}\equiv P(p^{\rm ren}_i)$ and 
$\Delta P\equiv P_{\rm obs}-P_{11}-P_{22}^{\rm SPT}-2P_{13}^{\rm ren}$. 
Combining Eqs.~\eqref{eq:Pren13}-\eqref{eq:C3ren}, we find the one-loop matter power spectrum, which can be used to test the DHOST theory. 
Contrary to GR, the two EFT parameters depend on the two renormalization scales. 
Thus, the running of EFT parameters are more complicated than that in GR.


\subsection{Screening mechanism}\label{sec: screening}


Here we discuss the relation between the apparent UV divergence and screening mechanism in the DHOST theory.
In modified gravity, the standard gravitational law should be reproduced by a screening mechanism which shields the propagation of the scalar field on small scales.
In the Horndeski and DHOST theories, non-linear derivative self-interactions of the scalar field shield its propagation, which is referred as ``k-mouflage/ kinetic screening"~\cite{Babichev:2009ee} or ``Vainshtein screening"~\cite{Vainshtein:1972sx}.
The screening mechanisms seem to be related to the UV divergence in the one-loop matter power spectrum as we argue below. However, to the best of our knowledge, no rigorous perturbative calculation
of LSS taking into account the screening mechanism has been done.
In Ref.~\cite{Fasiello:2017bot}, the authors have discussed the screening effects on the matter power spectrum in perturbative approaches, while the effects were phenomenologically introduced.

The great advantage of the EFTofLSS is that it automatically incorporates the screening effects in the counterterms as long as the cutoff $\Lambda$ corresponds to a longer scale than the screening scale $k_{\rm scr}$. This hierarchy $\Lambda<k_{\rm scr}$ usually holds. On one hand, $\Lambda$ is taken smaller than the non-linear scale, $k_{\rm nl}$, at which non-perturbative effects is significant $P_{\rm L}(k_{\rm nl})k^2_{\rm nl}\sim 1$.
On the other hand, for the screening mechanism to work, the non-linear effects have to be effective and the density fluctuation is typically larger than unity, which leads to $k_{\rm nl}<k_{\rm scr}$.
As a result, we have $\Lambda<k_{\rm nl}<k_{\rm scr}$. Therefore, in practice, we do not need to be concerned about the screening mechanisms in calculating the LSS observables in the EFTofLSS.

Still, the relation between the apparent UV divergence and the screening mechanism is of interest. Our speculative comments are in order. On scales where the screening mechanism works, the prediction of GR is restored and the $Q_{13}$ terms causing the UV divergence is supposed to vanish.  It may imply that the screening scale provides a built-in UV cutoff of the divergence in Eq.~\eqref{eq: P13MG} and the apparently divergent loop integral remains finite even without the EFTofLSS framework. Based on this observation, we suppose that at least one of the origins of the EFT counterterms is the screen mechanism. It would be fascinating to explore their relationship in the first principle. To this end, however, we need another approach which somehow deals with the full non-linear dynamics and the screening mechanism  because the EFTofLSS is agnostic to the origin of the counterterms.


\section{Conclusion}\label{sec: conclusion}


In this paper, we have extended the framework of the effective field theory of large scale structure (EFTofLSS) in general relativity (GR) to modified gravity. 
After that,  we applied our framework to the degenerate higher-order scalar-tensor (DHOST) theory.
It is shown that the ultraviolet (UV) divergence in the one-loop power spectrum is precisely canceled out thanks to the new counterterms, and we obtain a convergent matter power spectrum, which can be compared to observations for testing the DHOST theory.

In the procedure of the EFTofLSS in GR, the effective fluid is introduced to represent the effect of 
the non-linearity in the Euler equation. 
The effective fluid plays the role of counterterms and cancels out the UV sensitivity in the one-loop power spectrum.
In modified gravity theories, not only the Euler equation has non-linear terms, but also the Poisson equation is modified and have non-linearity.
We introduced the new three EFT terms in the field equation for $\Phi$~\eqref{eq: EFT terms}.
In the Horndeski theory and the DHOST theory, the quasi-static approximation is assumed for both background variables and perturbations.
Hence we need not consider the $c_1$ term.
The contribution of the $c_2$ term has a higher momentum dependence than that in GR in the one-loop power spectrum, while that of the $c_3$ term has the same momentum dependence as in GR.
We demonstrated the application of our framework to the DHOST theory, and we can require that the logarithmic divergence in the one-loop power spectrum is precisely canceled out.
In order to fix the new EFT parameters, it is necessary to use reference values of the power spectrum at two different wave numbers.

\acknowledgements

We would like to thank Matteo Fasiello, Ryo Saito, Atsushi Taruya, Chulmoon Yoo, and Zvonimir Vlah for fruitful discussions.
This work has been generated thanks to the online JGRG webinars. 
This work was supported in part by
JSPS KAKENHI Grant Nos. JP21H01080 (S.H.), JP18K13537, and JP20H05854 (T.F.).

\appendix

\section{SPT calculations in EFTofLSS in GR}\label{sec: EFT GR app}

Using the procedure of the EFTofLSS, we obtain the basic equations in Fourier space.
The Poisson equation does not change from the standard one. The continuity and Euler equations read
\begin{align}
    &\frac{1}{H}\dot\delta_l(t,{\bm p}) +\theta_l(t,{\bm p}) 
\notag\\    
    &= -\frac{1}{(2\pi)^3}\int^{\Lambda} d^3k~ \alpha({\bm p},{\bm k})\theta_l(t,{\bm p}-{\bm k})\delta_l(t,{\bm p}),\label{eq: continuityEFT}
\\
    &\frac{1}{H}\dot\theta_l(t,{\bm p}) +\left(2+\frac{\dot H}{H^2}\right)\theta_l(t,{\bm p}) +\frac{3}{2}\Omega_{\rm m} \delta_{l}(t,{\bm p})
\notag\\    
    &= -\frac{1}{(2\pi)^3}\int^{\Lambda} d^3k~ [\alpha_{s}({\bf p},{\bf k})-\gamma({\bf p},{\bf k})]\theta_l(t,{\bf p}-{\bf k})\theta_l(t,{\bf p})
\notag\\
    &\quad+c_{s}^2\frac{p^2}{a^2H^2}\delta_l(t,{\bf p})
    -c^2_{v}\frac{p^2}{a^2H^2}\theta_l(t,{\bf p})
    +\frac{p_{i}p_{j}}{a^2H^2}\Delta\tau^{ij}(t,{\bf p})
    ,\label{eq: EulerEFT}
\end{align}
where we contracted $\partial^{i}$ to Eq.~(\ref{eq: EulerR}) and used the definition of $\theta$, Eq.~(\ref{eq: theta}), and the Poisson Eq.~(\ref{eq: PoissonR}) with 
$c_{v}^{2} = c_{sv}^{2}+c_{bv}^{2}.$
We also have defined the quantities
\begin{align}
    \alpha({\bm k}_1,{\bm k}_2) &= 1 +\frac{{\bm k}_1\cdot{\bm k}_2}{k_1^2},
\\
    \alpha_s({\bm k}_1,{\bm k}_2) &= \frac{1}{2}[\alpha({\bm k}_1,{\bm k}_2)+\alpha({\bm k}_2,{\bm k}_1)],
\\
    \gamma({\bm k}_1,{\bm k}_2) &= 1 -\frac{({\bm k}_1\cdot{\bm k}_2)^2}{k_1^2k_2^2}.
\end{align}
The terms in the last line of Eq.~(\ref{eq: EulerEFT}) are additional terms from the EFT argument, while the other terms are the same as the SPT case, except for the index $l$ and the coarse-grained scale $\Lambda$ appearing at the upper limit of the integrals.
These EFT terms are treated as source terms in order to play a role of the counterterms.
Solving the system for the coarse-grained fields perturbatively, we obtain solutions by using the same procedure as the SPT way.  

Truncating non-linear terms in the fluid equations, the linearized equations are given by
\begin{align}
    &\frac{1}{H}\dot\delta_{l}^{(1)}(t,{\bm p})+\theta_{l}^{(1)}(t,{\bm p})=0,\label{eq: continuity1}
\\
    &\frac{1}{H}\dot\theta_{l}^{(1)}(t,{\bm p}) +\left(2+\frac{
    \dot H}{H^2}\right)\theta_{l}^{(1)}(t,{\bm p}) -\frac{3}{2}\Omega_{\rm m}\delta_l^{(1)}(t,{\bm p}) =0,\label{eq: Euler1}
\end{align}
Combining these equations, we obtain the linear evolution equation, 
\begin{align}
    \ddot\delta_{l}^{(1)}(t,{\bm p}) +2H\dot\delta_{l}^{(1)}(t,{\bm p}) -\frac{3}{2}H^2\Omega_{\rm m}\delta_{l}^{(1)}(t,{\bm p}) =0.\label{eq: evolution1}
\end{align}
This equation is a second-order ordinary differential equation such that we have growing and decaying solutions, $D_{+}(t)$ and $D_{-}(t)$ respectively.
Neglecting the decaying solution, we obtain the first-order solution~\eqref{eq: linear soln}.

At non-linear level, the convolutions and the EFT terms in the right hand side of Eqs.~(\ref{eq: continuityEFT}) and (\ref{eq: EulerEFT}) become relevant. 
In contrast to these fluid equations, the Poisson equation always remains in the original linear form even at non-linear level in GR.
Using the second-order equations and substituting the first-order solutions, the second-order evolution equation is given by
\begin{align}
    &\ddot\delta_{l}^{(2)}(t,{\bm p}) +2H\dot\delta_{l}^{(2)}(t,{\bm p}) -\frac{3}{2}H^2\Omega_{\rm m}\delta_{l}^{(2)}(t,{\bm p}) 
\notag\\    
    &= 
    \frac{H^2}{(2\pi)^3}\int^{\Lambda} d^3k\left[\frac{(a^2H^2D_{+}^3f)^{\dd}}{a^2H^2D_{+}}\alpha_{s}({\bf p},{\bf k}) -D_{+}^2f^2\gamma({\bf p},{\bf k})\right]
\notag\\    
    &\hspace{1.5cm}\times\delta_{L}({\bf p}-{\bf k})\delta_{L}({\bf p}).\label{eq: evolution2}
\end{align}
We assumed the density fluctuations do not initially have non-linearity, so that non-linear solutions are given by the inhomogeneous one.
The second-order solution is given by
\begin{align}
    \delta^{(2)}_{l}(t,{\bm p}) &=
    \frac{D_+^2(t)}{(2\pi)^3}\int^{\Lambda} d^3k~ F_2({\bm p},{\bm k})\delta_{l}({\bm p}-{\bm k})\delta_{l}({\bm k}),
\end{align}
where
\begin{align}
    F_2(t,{\bm p},{\bm k}) &= \kappa(t)\alpha_s({\bm p},{\bm k}) -\frac{2}{7}\lambda(t)\gamma({\bm p},{\bm k}),\label{eq: F2GR}
\notag\\
    \kappa(t) &:= \frac{1}{D_{+}^2}L\left[\frac{(a^2H^2D_{+}^3f)^{\dd}}{a^2D_{+}}\right],
\\
    \lambda(t) &:=
    -\frac{7}{2D_{+}^2}L\left[D_{+}^2H^2f^2\right],
\end{align}
and we defined the operator $L$ which acts on a function of time $s(t)$ as 
\begin{align}
    L[s] :=
    \int^{t}_{0}dT\frac{D_{+}(T)D_{-}(t) -D_{+}(t)D_{-}(T)}{D_{+}(T)\dot{D}_{-}(T) -\dot{D}_{+}(T)D_{-}(T)}s(T).
\end{align}
In the matter dominance (Einstein de-Sitter universe) in GR, we have $\kappa=\lambda=1$ exactly.
Even after the matter dominance, $f$ starts to evolve but the growth of $f$ is slower than $H$. So, $\kappa$ and $\lambda$ still are close to unity.
In this App.~\ref{sec: EFT GR app}, we regard $\kappa$ and $\lambda$ as unity for simplicity.
Using the same procedure at third order, we can also obtain the third-order solution as
Eq.~\eqref{eq: EFT3rd sol1} and
\begin{align}
    \delta^{(3)}_{\rm EFT2}(t ,{\bm p}) &= 
    -\int^{t}d\tau\, G(t,\tau)H^2k_{i}k_{j}\Delta\tau^{ij},
    \label{eq: delEFT2}
\\
    \delta^{(3)}_{\rm SPT}(t ,{\bm p}) &=  \frac{2D_+^4(t)}{(2\pi)^3}P_{L}(p)\int^{\Lambda} d^3k~ F_3({\bm p},{\bm k},-{\bm k})P_{L}(k),
    \label{eq: delSPT3}
\end{align}
with
\begin{align}
    F_3({\bm k}_1,{\bm k}_2,{\bm k}_3) &=
    \alpha\alpha({\bm k}_1,{\bm k}_2,{\bm k}_3) -\frac{4}{7}\alpha\gamma({\bm k}_1,{\bm k}_2,{\bm k}_3) 
\notag\\    
    &\quad -\frac{2}{21}\gamma\gamma({\bm k}_1,{\bm k}_2,{\bm k}_3) +\frac{1}{9}\xi_c({\bm k}_1,{\bm k}_2,{\bm k}_3).\label{eq: F3GR}
\end{align}
The explicit definitions of the third-order shape functions, $\alpha\alpha,~\alpha\gamma,~\gamma\gamma,~\xi_{\rm c}$, are written in App.~\ref{sec: 3rd shape funcs}.

Using the solutions up to third order, we obtain the one-loop corrections as Eqs.~\eqref{eq: 13SPT}, \eqref{eq: P13 EFT1}, and
\begin{align}
    P_{22}^{\rm SPT}(t,p) &= \frac{2D_+^4(t)}{(2\pi)^3}\int^{\Lambda} d^3k~ F^2_2({\bm k},{\bm p}-{\bm k})P_{L}(k)P_{L}(|{\bm p}-{\bm k}|),\label{eq: 22SPT}
\\
    P_{13}^{\rm EFT2}(t,p) &=
    -\frac{1}{(2\pi)^3}\int^{t}d\tau G(t,\tau)p_{i}p_{j}\langle\delta_l^{(1)}\Delta\tau^{ij}\rangle.
\end{align}

It the UV limit of the loop integrals, the SPT contributions behave as
\begin{align}
        P^{\rm SPT}_{13}(t,p) &\approx -\frac{61D_+^4(t)}{315(2\pi)^2}p^{2}P_{\rm L}(p)
\notag\\
    &\quad\quad \times\int^{\Lambda}_{p\ll k} dk ~P_{\rm L}(k)\left[1 +{\cal O}\left(\frac{p^2}{k^2}\right)\right]\label{eq: P13SPT}.
\\
    P^{\rm SPT}_{22}(t,p) &\approx \frac{135D_+^4(t)}{735(2\pi)^2}p^4\int^{\Lambda}_{p\ll k} dk~ \frac{P^2_{\rm L}(k)}{k^2} 
    \left[1 +{\cal O}\left(\frac{p^2}{k^2}\right)\right]\label{eq: P22SPT},
\end{align}
They should be compared to the external momentum dependence of the EFT contributions,
\begin{align}
    &P^{\rm EFT1}_{13}(t,p) \propto p^2P_{L}(p),
\\
    &P^{\rm EFT2}_{13}(t,p) \propto p^4,
\end{align}
where we used the result for the stochastic terms in Ref.~\cite{Peebles:1980co}.
These external momentum dependence are exactly the same.
Thus the EFT terms can work as counterterms in one-loop corrections.\footnote{Note that of course these effective terms can cancel out  leading parts of loop integrals. If one consider $\Lambda$-dependence of this renormalization method, one need to introduce sub-leading parts in an effective fluid, then one will see the cancellation of divergence in loop integrals at higher order.}

\section{3rd-order shape functions}\label{sec: 3rd shape funcs}

The 3rd-order solution for the density contrast has the several momentum dependence in GR and modified gravity. Here we explicitly write them.
The kernels that are generated by $\alpha_s$ and $\gamma$ are 
\begin{align}
    &\alpha\alpha({\bm k}_{1},{\bm k}_{2},{\bm k}_{3}) 
	=\frac{1}{3}\Bigl[
	\alpha_{s}({\bm k}_{1},{\bm k}_{2}+{\bm k}_{3})\alpha_{s}({\bm k}_{2},{\bm k}_{3}) 
    +2~{\rm perms.}
	\Bigr],\label{eq:alphaalpha def}
\\
	&\alpha\gamma({\bm k}_{1},{\bm k}_{2},{\bm k}_{3}) 
	=\frac{1}{3}\Bigl[
	\alpha_{s}({\bm k}_{1},{\bm k}_{2}+{\bm k}_{3})\gamma({\bm k}_{2},{\bm k}_{3}) 
    +2~{\rm perms.}
	\Bigr],
\\
	&\gamma\alpha ({\bm k}_1,{\bm k}_2,{\bm k}_3)
	=\frac{1}{3}\Bigl[
	\gamma ({\bm k}_1,{\bm k}_2+{\bm k}_3)\alpha_s({\bm k}_2,{\bm k}_3)
	+2~{\rm perms}
	\Bigr],
\\
	&\gamma\gamma({\bm k}_{1},{\bm k}_{2},{\bm k}_{3}) 
	=\frac{1}{3}\big[
	\gamma({\bm k}_{1},{\bm k}_{2}+{\bm k}_{3})\gamma({\bm k}_{2},{\bm k}_{3}) 
    +2~{\rm perms.}
    \big].\label{eq:gammagamma def}
\end{align}
In the DHOST theory, the new kernels appear, which are generated by the antisymmetric part of $\alpha$ as well as $\alpha_s$ and $\gamma$,
\begin{align}
	\alpha\alpha_\ominus({\bm k}_{1},{\bm k}_{2},{\bm k}_{3}) 
	&= \frac{1}{6}\Big\{
	\bigl[\alpha({\bm k}_{1},{\bm k}_{2}+{\bm k}_{3})-\alpha({\bm k}_{2}+{\bm k}_{3},{\bm k}_{1})
	\bigr]
\notag\\
	&\quad	\times\alpha_{s}({\bm k}_{2},{\bm k}_{3}) 
	+2~{\rm perms.}
	\Big\},\label{eq:alphaalpha- def}
\\
	\alpha\gamma_\ominus ({\bm k}_{1},{\bm k}_{2},{\bm k}_{3})
	&= \frac{1}{6}\Bigl\{
	\bigl[\alpha({\bm k}_{1},{\bm k}_{2}+{\bm k}_{3})-\alpha({\bm k}_{2}+{\bm k}_{3},{\bm k}_{1})
	\bigr]
\notag\\
	&\quad \times
	\gamma({\bm k}_{2},{\bm k}_{3}) 
	+2~{\rm perms.}
	\Bigr\}.\label{eq:alphagamma- def}
\end{align}
In addition to these six kernels, we need to consider two extra momentum dependence related to scalar non-linear self-interactions (see Ref.~\cite{Hirano:2020dom}), 
\begin{align}
    \xi({\bm k}_1,{\bm k}_2,{\bm k}_3) &= 
    1 -3\frac{({\bm k}_{2}\cdot{\bm k}_{3})^{2}}{k_{2}^{2}k_{3}^{2}} 
\notag\\    
    &\quad+2\frac{({\bm k}_{1}\cdot{\bm k}_{2})({\bm k}_{2}\cdot{\bm k}_{3})({\bm k}_{3}\cdot{\bm k}_{1})}{k_{1}^{2}k_{2}^{2}k_{3}^{2}},\label{eq:xi def}
\\
    \zeta({\bm k}_1,{\bm k}_2,{\bm k}_3) &= 
    \frac{({\bm k}_{2}\cdot{\bm k}_{3})^{2} }{k_{2}^{2}k_{3}^{2}}
    +2\frac{({\bm k}_{1}\cdot{\bm k}_{3})({\bm k}_{2}\cdot{\bm k}_{3})^{2}}{k_{1}^{2}k_{2}^{2}k_{3}^{2}}
\notag\\    
    &\quad +\frac{{\bm k}_{2}\cdot{\bm k}_{3}}{k_2^2}
    +\frac{({\bm k}_{1}\cdot{\bm k}_{2} +{\bm k}_{3}\cdot{\bm k}_{1})({\bm k}_{2}\cdot{\bm k}_{3})}{k_{1}^{2}k_{2}^{2}}.\label{eq:zeta def}
\end{align}
We used those functions symmetrized cyclically,
\begin{align}
    &\xi_{c}({\bm k}_{1},{\bm k}_{2},{\bm k}_{3}) =
    \frac{1}{3}\Bigl\{ \xi({\bm k}_{1},{\bm k}_{2},{\bm k}_{3}) +2~{\rm perms.}\Bigr\}
    ,\\
    &\zeta_{c}({\bm k}_{1},{\bm k}_{2},{\bm k}_{3}) =
    \frac{1}{3} \Bigl\{\zeta({\bm k}_{1},{\bm k}_{2},{\bm k}_{3}) +2~{\rm perms.}\Bigr\}
    .\label{eq:zeta_c def}
\end{align}

\section{Derivation of the schematic modified Poisson equation}\label{sec: schematic Poisson}

In this section, we derive the schematic modified Poisson equation, Eq.~(\ref{eq: modified Poisson}).
In scalar-tensor theories, the gravitational fields and scalar field are minimally coupled to matter in the Jordan frame.
In this situation, gravitational potentials and the scalar field fluctuations are coupled in equations of motion at both linear and non-linear levels while the fluid equations are still standard ones.
It is important to note that the equations of motion contain both non-linear scalar self-interactions and non-linear interactions between the scalar field and the gravitational potentials (see Eqs.~(7)--(9) in Ref.~\cite{Hirano:2020dom}).

We solve the system perturbatively.
Solving equations of motion for the gravitational potentials and the scalar field fluctuation algebraically at each order, substituting the solution for $\Phi$ into the fluid equations, we finally obtain the evolution equation for the density fluctuation.
Furthermore, to obtain the modified Poisson equation, we need to solve the evolution equation for the density fluctuation at each order, and substitute solutions into the Euler-Lagrange equation for $\Phi$, and sum up each order solutions.

In the DHOST theory, the first-order solution is given by~\cite{Hirano:2019nkz,Crisostomi:2019vhj,Lewandowski:2019txi}
\begin{align}
    \frac{\partial^2}{a^2H^2}\Phi^{(1)}(t,{\bm x}) &= \kappa_{\Phi}(t)\delta^{(1)}(t,{\rm x}) +\nu_\Phi(t)\frac{\dot \delta^{(1)}(t,{\bm x})}{H}
\notag\\
    &\quad +\mu_\Phi(t) \frac{\ddot\delta^{(1)}(t,{\bm x})}{H^2},
    \label{eq: 1st Phi}
\end{align}
where $\kappa_\Phi,~\nu_\Phi,~{\rm and }~\mu_{\Phi}$ are determined by solving linearlized equations of motion for gravitational potentials and scalar field.
The explicit forms are written in Ref.~\cite{Hirano:2019nkz,Hirano:2020dom}.
At second order, non-linear terms exists in equations of motion.
The solution is given by~\cite{Crisostomi:2019vhj,Lewandowski:2019txi,Hirano:2020dom}
\begin{align}
    \frac{\partial^2}{a^2H^2}\Phi^{(2)} &=\kappa_{\Phi}\delta^{(2)} +\nu_\Phi\frac{\dot \delta^{(2)}}{H}
    +\mu_\Phi \frac{\ddot\delta^{(2)}}{H^2}
\notag\\
    &\quad +\frac{\tau_{\Phi,\alpha}^{(2)}}{D_{+}^2}\left(\delta^{(1)}\delta^{(1)} +\frac{\partial_i\delta^{(1)}}{\partial^2}\partial_i\delta^{(1)}\right)
\notag\\
    &\quad
    +\frac{\tau_{\Phi,\gamma}^{(2)}}{D_{+}^2}\left\{\delta^{(1)}\delta^{(1)} -\left(\frac{\partial_i\partial_j\delta^{(1)}}{\partial^2}\right)^2\right\}.\label{eq: 2nd Phi}
\end{align}
where the coefficients, $\tau^{(2)}_{\Phi,i}~(i=\alpha,\gamma)$, are written by the grow rate and background variables in each modified gravity model (the explicit forms are given by Eq.~(B9) in Ref.~\cite{Hirano:2020dom}). 
We can rewrite $\dot\delta^{(1)}$ and $\ddot\delta^{(1)}$ to $\delta^{(1)}$ in the right hand side of Eq.~(\ref{eq: 1st Phi}) by using the linear solution for the density fluctuation (\ref{eq: linear soln}).
Substituting Eq.~(\ref{eq: 1st Phi}) into Eq.~(\ref{eq: 2nd Phi}), we obtain
\begin{align}
    \frac{\partial^2}{a^2H^2}\Phi^{(2)} &=\kappa_{\Phi}\delta^{(2)} +\nu_\Phi\frac{\dot \delta^{(2)}}{H}
    +\mu_\Phi \frac{\ddot\delta^{(2)}}{H^2}
\notag\\
    &\quad -\frac{1}{a^4H^4}\tilde\tau_{\Phi,\alpha}^{(2)}\left\{\left(\partial^2\Phi^{(1)}\right)^2 +\partial_i\Phi^{(1)}\partial_i\partial^2\Phi^{(1)}\right\}
\notag\\
    &\quad
    -\frac{1}{a^4H^4}\tilde\tau_{\Phi,\gamma}^{(2)}\left\{\left(\partial^2\Phi^{(1)}\right)^2 -\left(\partial_i\partial_j\Phi^{(1)}\right)^2\right\},\label{eq: 2nd Phi2}
\end{align}
where
\begin{align}
    \tilde\tau_{\Phi, i}^{(2)} =-\frac{\tau_{\Phi, i}^{(2)}/D_{+}^2}{\kappa_\Phi 
    +\nu_\Phi f 
    +\mu_\Phi\left[(fH)^{\dd}/H^2 +f^2\right]}
    ~(i=\alpha,\gamma).
\end{align}
Using the same procedure, we can also obtain higher-order solutions.
Summing up these perturbative solutions, we obtain the schematic modified Poisson equation~\eqref{eq: modified Poisson}, $i.~e.~$,
\begin{align}
  \frac{1}{a^2H^2}\partial^2\Phi
  +\frac{1}{a^4H^4}{\cal T}_{\rm NL}
  +\cdots
  = \kappa_{\Phi}\delta+\nu_\Phi\frac{\dot \delta}{H}
    +\mu_\Phi \frac{\ddot\delta}{H^2},
\end{align}
with
\begin{align}
    {\cal T}_{\rm NL} &=
 \tilde\tau^{(2)}_{\Phi,\alpha} \left\{(\partial^2\Phi)^2+\partial_{i}\Phi\partial_{i}\partial^2\Phi\right\}
\notag\\  
  &\quad
   +\tilde\tau^{(2)}_{\Phi,\gamma} \left\{(\partial^2\Phi)^2-(\partial_{i}\partial_{j}\Phi)^2\right\}.
\end{align}
The ellipses denotes higher-order terms for $\Phi$, namely $\mathcal{O}(\Phi^3)$,
because otherwise another quadratic term would have appeared in Eq.~\eqref{eq: 2nd Phi2}.

\bibliography{Refs}

\end{document}